\documentclass[prb,showpacs,superscriptaddress]{revtex4}
\usepackage{graphicx,amssymb,amsmath,color}
\begin{document}
\title{Rabi Oscillations in Landau Quantized Graphene}
\author{Bal\'azs D\'ora}
\email{dora@pks.mpg.de}
\affiliation{Max-Planck-Institut f\"ur Physik Komplexer Systeme, N\"othnitzer Str. 38, 01187 Dresden, Germany}
\author{Klaus Ziegler}
\affiliation{Institut f\"ur Physik, Universit\"at Augsburg, D-86135 Augsburg, Germany}
\author{Peter Thalmeier}
\affiliation{Max-Planck-Institut f\"ur Chemische Physik fester Stoffe, 01187 Dresden, Germany}
\author{Masaaki Nakamura}
\affiliation{Max-Planck-Institut f\"ur Physik Komplexer Systeme, N\"othnitzer Str. 38, 01187 Dresden, Germany}
\date{\today}

\begin{abstract}
The canonical model of quantum optics,
the Jaynes-Cummings Hamiltonian describes a two level
atom in a cavity interacting with electromagnetic field. 
Graphene, a condensed matter system, possesses low energy excitations obeying to the Dirac 
equation, and mimics the physics of quantum electrodynamics.
These two seemingly unrelated fields turn out to be closely related to each other.
We demonstrate that Rabi oscillations, corresponding to the excitations of the atom in the
former case are observable in the optical response of the latter in quantizing magnetic field, providing us
with a transparent picture about the structure of optical transitions in
graphene. While the longitudinal conductivity reveals chaotic Rabi
oscillations, the Hall component measures coherent ones.
This opens up the exciting possibility of investigating a microscopic model
of a few quantum objects in a macroscopic experiment of a bulk material
with tunable parameters.  
\end{abstract}


\maketitle

\begin{figure}[h!]
\includegraphics[width=2.6cm,height=5cm,angle=90]{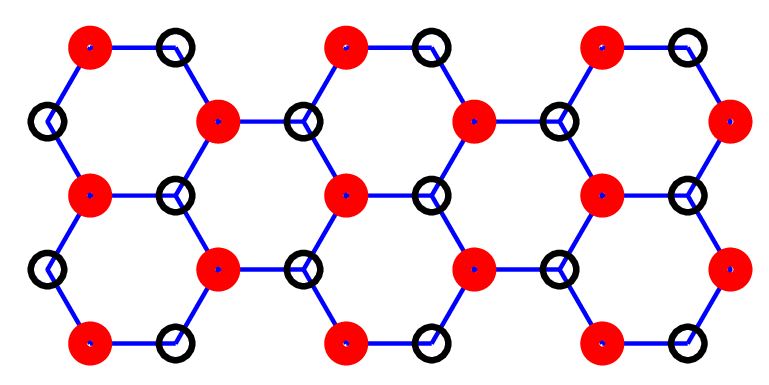}
\caption{(Color online) A small fragment of the honeycomb lattice is shown, the filled red and empty black circles denote the two 
sublattices.}
\label{subhoney}
\end{figure}

Graphene, a single sheet of carbon atoms in a honeycomb lattice has attracted enormous interest 
recently\cite{novoselov1}. 
Its quasiparticle states are pseudospinors, where the spinor
structure is a consequence of the two-fold sublattice structure of the honeycomb lattice (Fig. \ref{subhoney}). They obey a 
two-dimensional Dirac equation, whose speed of light is
replaced by the Fermi velocity (being 1/300th the speed of light).
This implies a number of striking properties, including the unconventional quantum Hall 
effect\cite{novoselov2}, Klein tunneling\cite{klein} and Zitterbewegung\cite{cserti,schliemanngraphene,rusin} due to
particle-hole excitations.
In an applied magnetic field, perpendicular to the carbon sheet, the formation of Landau 
levels $E_n$ with an unusual dependence
$E_n\sim\sqrt{n}$ on the Landau level index $n$ was predicted, and also observed 
experimentally\cite{andrei,jiang}.

Coupling of a (pseudo)spinor to an external quantum field is a common feature in
quantum optics, which deals with the interaction of electromagnetic field and matter. 
In this context, one of the simplest, fully quantum mechanical model is the 
Jaynes-Cummings (JC) Hamiltonian\cite{shore}
\begin{equation}
H_{JC}= V(a^+\sigma^-+\sigma^+a)+\Delta\sigma_z \ ,
\label{ham00}
\end{equation}
which describes a single two-state atom, represented by the Pauli matrices, that is
interacting with a (single-mode quantized) electromagnetic field\cite{barnett}.
$a^+$ ($a$) are the photon creation (annihilation) operators, and $V$ is the coupling strength
between the atom and the electromagnetic field.
The interaction leads to a periodic exchange of energy between the electromagnetic field 
and the two-level system, known as Rabi oscillations. This effect can be interpreted as a 
periodic change between absorption and stimulated emission of photons. It is simply
a consequence of the fact that the atom is not in an eigenstate after the absorption
of $n$ photons. A very similar situation occurs for the Landau levels of graphene. 
These levels play the role of the atomic levels in quantum optics due to photon absorption
and the sublattice structure of graphene provides the two-fold degeneracy of the two levels
of the atom. In terms of the Hamiltonian (\ref{ham00}), the photon operators $a^+$ and
$a$ are operators that are acting on Landau levels\cite{peresalap}:
$a^+=\pi^+/\sqrt{2eB}$ and $a=\pi^-/\sqrt{2eB}$, where 
$\pi^\pm=\pi_x\pm\pi_y$ with ${\bf \pi=p}+e\bf A$ are the (Peierls
substituted) momentum operator in the vector potential $\bf A$.
Formally, they act as creation and annihilation operators of a
harmonic oscillator and obey a bosonic commutation relation $[\pi^-,\pi^+]=2eB$.
Moreover, the coupling constant $V$ depends on the magnetic field $B$ 
and the Fermi velocity in graphene $v_F=10^6$~m/s through $V=v_F\sqrt{2eB}$.
Finally, $\Delta$ represents a possible excitonic gap\cite{khveshchenko} or a substrate 
induced bandgap\cite{bandgap} in epitaxial graphene.

The pseudospinor state in the $n$th Landau level 
is not an eigenstate of the quasiparticle Hamiltonian of graphene.  
If we prepare an initial quasiparticle state of graphene in a certain Landau 
level by switching on an external magnetic field, the quasiparticles will go from the initial
non-equilibrium state to other Landau levels. Depending on the damping due to
(impurity) scattering, the quasiparticles oscillate between different Landau levels until
they reach their equilibrium state, which is a superposition of Landau levels. Recent
experiments on high-mobility samples of graphene have indicated that damping due to 
impurity scattering can be negligible on experimentally relevant time scales \cite{geim,kim}.  
 
In this Letter we shall discuss the non-equilibrium current dynamics of high-mobility 
graphene. Quantum states are prepared by a short electric pulse and the subsequent 
current dynamics is controlled by oscillations between Landau levels.
 
The Hamiltonian $H_{JC}$
can be diagonalized  for Landau levels, and the resulting eigenvalues are obtained as
$E_{n\alpha}=\alpha\sqrt{\Delta^2+V^2(n+1)}$,
where $n=0,$ 1, 2\dots non-negative integer, $\alpha=\pm$. 
In the non-relativistic limit ($\Delta\gg V$), the usual Landau level spectrum is obtained as 
$\alpha(\Delta+\omega_c(n+1))$ with the
cyclotron frequency of massive (Dirac) fermions $\omega_c=v_F^2eB/\Delta$.
In addition, there is a special eigenstate, stemming from the Landau level at the Dirac point 
with
$E^*=-\Delta$, which formally corresponds to $n=-1$ and $\alpha=-1$.
Having determined the spectrum of the Hamiltonian, we turn to the
investigation of the current correlations. Since $\sigma_{x(y)}$ is the current density
in $x$ ($y$) direction due to the equation of motion $j_{x(y)}=i[H,x(y)]=v_F\sigma_{x(y)}$, 
we can calculate the dynamical
correlation function $C_{xx}(t)=\langle\sigma_x(t)\sigma_x(0)\rangle$ and
$C_{xy}=\langle \sigma_x(t)\sigma_y(0)\rangle$ (symmetric and antisymmetric dipole-dipole correlator).
In quantum optics, these describe the transitions
between the two atomic states, and tell us about the spectrum of Rabi
oscillations\cite{agarwal}.
On the other hand,  $C_{xx}(t)$ and $C_{xy}(t)$ play the role of the longitudinal and Hall 
current-current correlation
functions in the Dirac case, and leads eventually to the optical 
conductivity of graphene\cite{opticalgeim,zqli}. 
Therefore, we expect the well-known Rabi oscillations of
quantum optics characterizing the excitations of the atom to be observable in the response functions of Landau 
quantized Dirac fermions\cite{brune}.

We start with the general correlator 
$C(\varphi,t)=\langle\sigma_x(t)(\cos(\varphi)\sigma_x+\sin(\varphi)\sigma_y)\rangle$, 
evaluated using the (structureless) bosonic $a$ operators. This defines both $C_{xx}(t)=C(0,t)$ and
$C_{xy}(t)=C(\pi/2,t)$, and reads as 
\begin{gather}
C(\varphi,t)=
\sum_{n\geqslant 0\alpha\gamma s=\pm} g_cf(E_{n\alpha})\exp(i(E_{n\alpha}+E_{n-s\gamma})t+is\varphi)
P_{n\alpha\rightarrow 
n-s\gamma}+
g_cf(E^*)\sum_{\gamma=\pm}\exp(i(E_{0\gamma}+E^*)t-i\varphi)P_{*\rightarrow 0\gamma},
\label{cxxtcorr}
\end{gather}
where $f(E)=1/(\exp((E-\mu)/T)+1)$ is the Fermi function,
$g_c=N_fA_ceB/2\pi$ is the degeneracy of the Landau levels and spins, $A_c$ is
the area of the unit cell, to be taken as unity in the Dirac approach,
$N_f=2$ stands for the spin degeneracy.
From this, it follows immediately that a $E_{n\alpha}$ 
Landau level with $n>0$ and given $\alpha$ possesses 4 
possible optical transitions to the adjacent levels as 
$E_{n\pm 1\pm\alpha}$ (on the same side and on the other 
side of the Dirac cone), the $n=0$ level 3 transitions to 
$E_{1\pm\alpha}$ and $E^*$ and the $E^*$ level two transitions to $E_{0\pm}$.
The non-zero transition matrix elements  are given by
\begin{gather}
P_{n\alpha\rightarrow n-s\gamma}=\frac 14 \left(1+\frac{s\Delta}{E_{n\alpha}}\right)
\left(1+\frac{s\Delta}{E_{n-s\gamma}}\right) \textmd{ for }
n\geqslant 0, \textmd{ and } 
P_{*\rightarrow 0\gamma}=\frac 12 \left(1-\frac{\Delta}{E_{0\gamma}}\right),
\end{gather}
which satisfy $\sum_{m\gamma}P_{n\alpha\rightarrow m\gamma}=1$, and 
agree with the transition probabilities for Rabi oscillations of
atoms induced by external electromagnetic field. These 
approach $1/4$ in the classical limit (of bosons) $n\rightarrow
\infty$, in which case the field contains many bosons, whose quantum
character can then be neglected\cite{bermudez}. Interestingly, the $\Delta=0$ limit
yields the classical matrix elements for any $n\geqslant 0$.
However, the $E^*$ level never reaches the classical limit, and is
responsible for the anomalous optical properties of graphene in magnetic field\cite{ando2}.
The evaluation of the microwave conductivities of graphene follows readily from Eq. \eqref{cxxtcorr},
and the response function is
$\chi_{xa}(t)=-\Theta(t)2e^2v_F^2\textmd{Im}C_{xa}(t)$, which
yields to the optical ($a=x$) and microwave Hall ($a=y$) conductivity as
$\sigma_{xa}(\omega)=\tilde\chi_{xa}(\omega)/i\omega$ upon Fourier transformation. 

Eq. \eqref{cxxtcorr}, together with the relation to the JC Hamiltonian
provides us with a particularly simple picture about the optical selection
rules and transitions by relating them to the Rabi oscillations.
Therefore, the optical conductivity by varying the frequency sweeps through
all possible transitions, and measures the frequency of the Rabi
oscillations, with quantum or classical character.
This provides a unique opportunity to investigate a basic phenomenon of
quantum electrodynamics in a condensed matter experiment.
By changing the external magnetic field applied to graphene, the coupling between the atom and
electromagnetic field in the JC model can be tuned
continuously, facilitating the exploration of various regimes, the quantum
to classical crossover.

\begin{figure}[h!]
\includegraphics[width=8cm,height=6cm]{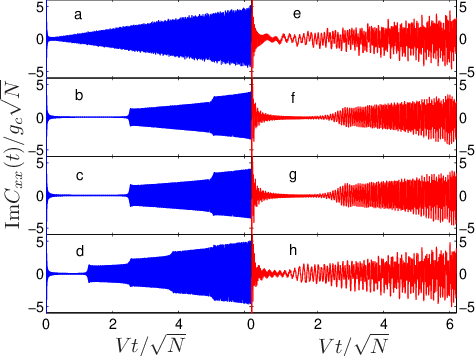}
\caption{(Color online) The real time evolution of $C_{xx}(t)$ is shown at $T=0$, taking both valley and spin 
degeneracies into 
account. We introduced a cutoff $D$, and the 
number of levels is measured as $D=V\sqrt{N}$, corresponding to different magnetic field strengths. The left/right 
panels show $N=10000$ (blue)/$N=100$ (red) with 
$D=V\sqrt{N}$ for $(\mu,\Delta)/V\sqrt{N}=(0,0)$ [a/e], (0.4,0) [b/f],
(0.4,0.2) [c/g], and (0,0.2) [d/h]. These structures correspond to thermal field induced random oscillation in quantum 
optics\cite{knight}.}
\label{cxxt1e4_1e2}
\end{figure}

\begin{figure}[h!]
\includegraphics[width=8cm,height=2.2cm]{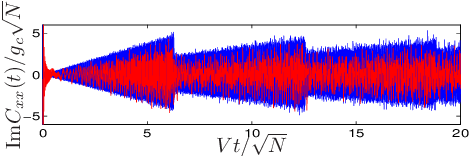}
\caption{(Color online) The real time evolution of $C_{xx}(t)$ is shown for long times for $T=0$, $(\mu,\Delta)=(0,0)$. 
We introduced a cutoff $D$, and the
number of levels is measured as $D=V\sqrt{N}$, corresponding to different magnetic field strengths with 
$N=10000$ (blue) and $N=100$ (red). Collapse and revival shows up with time similarly to the thermal field 
Jaynes-Cummings  model. The revivals gradually get wider and overlap. The presence of thermal revivals are related to 
the finite average boson number in the Jaynes-Cummings model, which translate to a finite cutoff in the Dirac case.
As opposed to Fig. \ref{cxxt1e4_1e2}, these revivals at long times are caused by the finite cutoff.}
\label{cxxtlong}
\end{figure}

In quantum optics, one has the freedom to prepare the initial state
of both the atom and the electromagnetic field\cite{shore,barnett}. The atom is usually prepared in its
excited state, and the field is prepared in a number state or in a coherent
state\cite{brune,rempe}. Then one can study the time evolution of the atomic population, which
exhibits Rabi oscillations, when jumping between the ground and excited
state, causing collapse and revival phenomenon.
However, qualitatively different behaviour describes chaotic or thermal
fields\cite{knight}. 
Quiescent periods  and interfering revivals  are also
present, but the resulting pattern of oscillations follows an apparently
random evolution.
A coherent state, which is the most classical single mode quantum state, is
strongly peaked around the average boson number $\bar n$. Therefore, Rabi
oscillations mainly involve frequencies around $E_{\bar n\alpha}$, and the
collapse time is independent of the field strength, i.e. $\bar n$.
As opposed to this, a thermal or chaotic field has a broad, monotonically
decreasing distribution
of boson states, and can be represented as a mixture of coherent states
with a Gaussian distribution of the mean values. As a result, the very wide
range of boson numbers gives such a broad distribution of Rabi frequencies
that almost no trace of coherent oscillations remains after ensemble
averaging, and the resulting initial collapse time depends on the field
strength $\bar n$. The revival times depend on $\bar n$ for both initial conditions.

For Graphene, an arbitrary preparation
of the initial states is not accessible, but requires thermal, ensemble averaging.
In this respect, it is closer to the second type of thermal initial condition for
the JC model. The average boson number in the JC model
corresponds to the total number of fermions in graphene,
determined by the chemical potential and the cutoff. This can be introduced
by the energy scale $D=V\sqrt{N+1}$, above which we neglect all states
(with $n>N$). We mention that the inclusion of a cutoff is required to obtain correctly the f-sum rule for 
graphene\cite{sabio}.

Using this prescription, we investigate the real time evolution of the
longitudinal current-current correlation function based on Eq. \eqref{cxxtcorr}.
The results are shown in Fig. \ref{cxxt1e4_1e2}, including valley and spin degeneracies.
Similarly to observations in quantum optics\cite{knight,shore}, the initial
collapse is followed by a revival of oscillation, which are also sensitive
to the presence of finite $\mu$ and $\Delta$. They both enlarge the quiescent
period after the short time collapse, and cause additional step-like
structures in the envelope of oscillations with a period of max$(\mu,\Delta)\pi/v_F^2eB$. For longer times, collapse and
revival is observable in Fig. \ref{cxxtlong}, which gradually become wider
and overlap. This revival time depends on the value of the cutoff like $2\pi \sqrt{N+1} /V=\pi D/v_F^2eB$, 
as is apparent from the figure, and is controllable by the magnetic field.

\begin{figure}[h!]
\includegraphics[width=8cm,height=5cm]{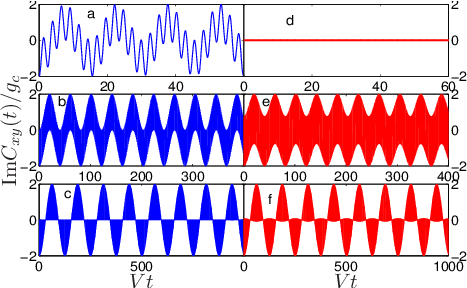}
\caption{(Color online) The real time evolution of $C_{xy}(t)$ is shown for $T=0$. The explicit value of the number of 
levels 
($N$)
does not influence the resulting pattern. The chemical potential varied as $\mu/V=1.2$ [a/d], 3.2 [b/e]  and 10 [c/f] with $\Delta=0$ 
(left
panel, blue) and $\Delta=2V$ (right panel, red), and the frequency of the envelope function is $v_F^2eB/\mu$, the 
cyclotron frequency of massless Dirac fermions. For $T=0$ and $\Delta>\mu$, Im$C_{xy}(t)=0$. Note the different 
horizontal scales!}
\label{imhallt}
\end{figure}

The Hall response evolves differently. In the DC limit, by taking
both valley and spin degeneracies into account, it produces the unconventional quantum Hall steps 
as a function of $\mu$. In connection with the
JC model, these steps, occurring when $\mu$ coincides with a
Landau level energy, correspond to the bare Rabi frequencies, which
can be revealed in a static, DC experiment, but still gaining information
about the dynamics of the system.
The microwave Hall conductivity obeys to the same selection rules and possesses the same 
transition matrix elements as the longitudinal conductivity. However,
the extra phase factor in Eq. \eqref{cxxtcorr} is responsible for a different
behaviour.

The time evolution of the Hall correlator is shown in Fig. \ref{imhallt}, and turns out to be independent of the 
applied cutoff scheme, hence being universal. It exhibits coherent Rabi oscillations, which vanish at the Dirac 
point (this is equivalent to the statement, that the Hall conductivity is zero exactly at the Dirac point). Upon 
increasing $\mu$, oscillations shows up with beating property, observed in quantum optics as well\cite{shore}. 
The characteristic frequency of
the envelope of the oscillations for $\mu\gg V,\Delta$ is $V^2/2\mu=v_F^2 eB/\mu$, which is the
cyclotron frequency of massless Dirac fermions\cite{novoselov2}. Even though the spectrum is 
linear, the finite chemical potential provides us with an energy scale for the cyclotron mass.
Only the energy levels with $|E_{n\alpha}|<\mu$ contribute to the Hall response at $T=0$.
Therefore, for $\Delta>\mu$, Im$C_{xy}(t)=0$ and no oscillations are
present. This means, that only a
finite, narrow range of frequencies determine the Rabi oscillations, 
similarly to the coherent field case in quantum optics, which
lead to beating and non-chaotic collapse and revival.
Thus, although the initial field state is always a thermal one in graphene, both
coherent and chaotic Rabi oscillation can be observed in different
quantities, together with collapse and revival.
Thus, the correlation function $C(\varphi,t)$ measures in
principle the crossover from thermal to coherent behaviour with changing
$\varphi$, and provides us with the unique opportunity to observe the
crossover by not changing the initial field, but by measuring a different
component of the current.

Time-resolved current-voltage measurements can reveal the real-time dependence of correlation functions, similarly to 
the current-voltage characteristics of Josephson junctions\cite{fenton}.
By applying a sharp current or electric field pulse as $E=E_0\delta(t)$, the resulting current through the 
sample parallel or perpendicular to $E_0$ after the initial pulse is directly related to the above correlation 
functions. More precisely, 
within linear response theory, it follows as
$\langle j_a(t)\rangle=-2e^2v_F^2E_0\int_0^t\text{d}t'\text{Im}C_{xa}(t')$ with $a=x$ or $y$. This opens up the 
possibility to reconstruct the time-dependence of Im$C_{xa}(t')$, or to obtain the frequency dependent longitudinal and 
Hall responses.
Another method invokes the femto/atto-second laser pulse technique. After shining the sample with a short laser pulse, 
the measurement of transmittance or reflectance in time is determined by the $C(t)$ correlation functions.
The most conventional way to deduce these correlation functions is provided through optical conductivity or 
current fluctuation 
measurement. Via the fluctuation dissipation theorem, they contain the same information, and upon Fourier transforming 
from frequency space to get the real time dependence, one is expected to be able to observe the presence of thermal Rabi 
oscillations. Similar measurements have already been carried out without magnetic field\cite{opticalgeim,zqli}, 
by exploiting the tunability of the carrier concentration with gate voltage. 

The above calculations can easily be extended to other correlation functions
such as thermal conductivities.
We also may speculate that various extensions to single layer graphene such as
bi- and multilayer structures\cite{mccannbilayer} (with spin-orbit coupling) can be mapped onto different 
multimode, multiatom and non-linear versions of the JC model.

We have shown that the equivalence of the Hamiltonians of graphene in magnetic field and of the JC model
influences their correlation functions as well, causing both thermal and coherent Rabi oscillation in the electric 
response of graphene.
Finally we speculate that Rabi oscillations and
Zitterbewegung are two closely related phenomena named differently in
different fields of physics, both arising from the coupling of positive and negative energy states.

This work was supported by the Hungarian Scientific Research Fund under grant number K72613 and in part
by the Swedish Research Council.

\bibliographystyle{apsrev}
\bibliography{refgraph}
\end{document}